\documentclass{cccg25}
\usepackage{graphicx,amssymb,amsmath}
\usepackage[T1]{fontenc}
\usepackage{xcolor}
\usepackage{hyperref}
\usepackage{cleveref}
\usepackage[all]{hypcap}
\usepackage{import}

\usepackage{varwidth}
\usepackage{usebib}
\ifcsname newbibfield\endcsname
    \newbibfield{author}
    \bibinput{article}
\fi
\newtheorem{definition}[theorem]{Definition}

\crefname{conj}{conjecture}{conjectures}
\crefname{obs}{observation}{observations}
\crefname{prop}{proposition}{propositions}
\crefname{lemma}{lemma}{lemmas}

\newenvironment{moral}{\quote}{\endquote}

\usepackage{oplotsymbl}
\newcommand\QQ{\mathbb{Q}}
\newcommand\CC{\mathbb{C}}
\newcommand\RR{\mathbb{R}}
\newcommand\Frac{\operatorname{Frac}}

\DeclareMathOperator\glue{glue}
\DeclareMathOperator\shrink{shrink}
\newcommand\Cccv {C_{\mathsf{ccv}}}
\newcommand\Ccvx {C_{\mathsf{cvx}}}
\newcommand\Cpoly{C_{\mathsf{poly}}}
\newcommand\Ctwin{C_{\mathsf{twin}}}
\begin{document}
\newcommand\CH{\operatorname{CH}}
\newcommand\absCH[1]{\lvert\CH(#1)\rvert}

\title{Counting Number of Triangulations of Point Sets: Reinterpreting and Generalizing the Triangulation Polynomials}
\author{Hong Duc Bui}
\maketitle

\begin{abstract}
    We describe a framework that unifies the two types of polynomials introduced
    in \cite{bacher2010,rutschmann2023}
    to analyze the number of triangulations of point sets.
    Using this insight, we generalize the triangulation polynomials of chains to a wider
    class of near-edges, enabling efficient computation of the number of triangulations
    of certain families of point sets.
    We use the framework to try to improve the result in \cite{rutschmann2023} without success,
    suggesting that their result is close to optimal.
\end{abstract}


\section{Introduction}

We are concerned with the problem of counting the number of possible ways to \emph{triangulate} a given point set.

\begin{figure}
    \centering
    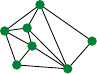
    \caption{Example of a triangulated point set.}
    \label{fig:sample_triangulated_point_set}
\end{figure}

See \cref{fig:sample_triangulated_point_set} for an illustration of a triangulation.
More formally, given a set of distinct points $P$ on the plane with no three points collinear,
an edge of $P$ is a line segment connecting two points in $P$,
and a triangulation of $P$ is a maximal set of non-intersecting edges.

\begin{figure}
    \centering
    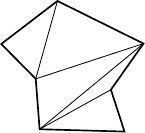
    \caption{Example of a triangulated polygon.}
    \label{fig:sample_triangulated_polygon}
\end{figure}

Sometimes, we are also concerned with the problem of counting the number of possible ways to \emph{triangulate} a given \emph{polygon} instead.
Here, instead of a point set we are given a non-self-intersecting polygon $P = P_1 P_2 \dots P_n$,
and a triangulation is a maximal set of non-intersecting diagonals of $P$.
We still assume no three points within the vertices of $P$ are collinear.
It can be shown that the triangulation consists of exactly $n-3$ diagonals, where $n$ is the number of sides of $P$.
See \cref{fig:sample_triangulated_polygon} for an illustration, note that unlike the case of triangulating a point set, edges lying outside $P$ is not considered.

We ask the following question:
\begin{quote}
    Let $f(n)$ be the maximum number of triangulations of a point set with $n$ points.
    What is the rate of growth of $f(n)$?
\end{quote}

We know that $f(n) \in O(30^n)$ thanks to a series to successive improvement
\cite{smith1989,denny1997,seidel1998,santos2003,sharir2006,sharir2011}.

Thanks to another series of successive improvements on the opposite direction
\cite{garcia2000,aichholzer2007,dumitrescu2013,rutschmann2023},
it is known that $f(n) \in \Omega(9.08^n)$.
The method of proving a lower bound $f(n) \in \Omega(c^n)$ is to provide an explicit family of point sets
and prove that it has at least $\Omega(c^n)$ triangulations.
Usually, proving the number of triangulations is the hard part.
Part of the reasons for the difficulty is that
the known methods for counting number of triangulations
of a point set takes exponential time \cite{alvarez2013,marx2016},
but more importantly, there is usually not enough structure on the collection of triangulations to
effectively compute the number of triangulations in closed-form.

This is where the technique in \cite{rutschmann2023} proves useful:
it defines a family of point sets, called \emph{chains}, where to each chain $C$
a \emph{upper triangulation polynomial}
$T_C (x)$ is associated. This polynomial can be computed in quadratic time
with respect to the number of points on the chain;
furthermore, the algorithm to compute the triangulation polynomials
can be easily modified to compute the total number of triangulations.

In this article, we define a generalization, \emph{joint triangulation polynomials}
$a_A^{yu}(y, u)$ which is a bivariate polynomial, and defined for all point sets $A$
that is a \emph{near-edge}.

This article is organized as follows.
In \cref{sec_motivation}, an exposition of existing ideas is explained
in order to motivate the definition of the maps $\mathcal T$ and $\mathcal M$.
In \cref{sec_formal_definition}, we formally define a near-edge and related concepts.
In \cref{sec_algebraic_definitions}, we formally define the joint triangulation polynomials
associated to each near-edge, which is to be used in \cref{sec_application_count_triangulations}
to count the number of triangulations of a near-edge, as well as a point set with near-edges glued to
its sides.
In \cref{algebraic_MT}, we further develop the algebraic theory,
which is used in \cref{another_analysis_polytwin} to simplify the statement of a certain theorem
in \cite{rutschmann2023}.
In \cref{sec_numerical_experiment}, we conjecture a similar statement for near-edge,
and describe experiments conducted to attempt to improve the bound $\Omega(9.08^n)$ described above.
The data obtained from running the experiments are included in \cref{sec_numerical_experiment_result}.

\section{Motivation}
\label{sec_motivation}

\subsection{Count Triangulations by Inclusion-Exclusion: Example}

We start with an exposition of \cite[Proposition 1]{hurtado1997}.

For an arbitrary polygon $P$, let $T(P)$ be the number of triangulations of $P$.
We may also draw the polygon itself in the place of the polygon,
for example $T(\triangle) = 1$ (the polygon is literally a triangle).
We also have for example $T(\square) = 2$, $T(\pentago) = 5$, $T(\hexago) = 14$, this is just the Catalan numbers.

Formally, define the $n$-th Catalan number to be $C_n = \frac{1}{2n+1} \binom{2n+1}{n}$,
then when $P$ is a convex polygon with $n$ edges, $T(P) = C_{n-2}$.


\begin{prop}
    \label{prop_simple_pie}
    \[ 
\begingroup%
  \makeatletter%
  \providecommand\color[2][]{%
    \errmessage{(Inkscape) Color is used for the text in Inkscape, but the package 'color.sty' is not loaded}%
    \renewcommand\color[2][]{}%
  }%
  \providecommand\transparent[1]{%
    \errmessage{(Inkscape) Transparency is used (non-zero) for the text in Inkscape, but the package 'transparent.sty' is not loaded}%
    \renewcommand\transparent[1]{}%
  }%
  \providecommand\rotatebox[2]{#2}%
  \newcommand*\fsize{\dimexpr\f@size pt\relax}%
  \newcommand*\lineheight[1]{\fontsize{\fsize}{#1\fsize}\selectfont}%
  \ifx\svgwidth\undefined%
    \setlength{\unitlength}{239.17201996bp}%
    \ifx\svgscale\undefined%
      \relax%
    \else%
      \setlength{\unitlength}{\unitlength * \real{\svgscale}}%
    \fi%
  \else%
    \setlength{\unitlength}{\svgwidth}%
  \fi%
  \global\let\svgwidth\undefined%
  \global\let\svgscale\undefined%
  \makeatother%
  \begin{picture}(1,0.10850109)%
    \lineheight{1}%
    \setlength\tabcolsep{0pt}%
    \put(0,0){\includegraphics[width=\unitlength,page=1]{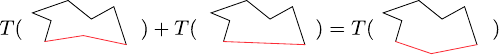}}%
  \end{picture}%
\endgroup%
 \]
\end{prop}

Notice that the \emph{only} difference between the $3$ polygons depicted in the equation is within the edges colored red.

\begin{figure}
    \centering
    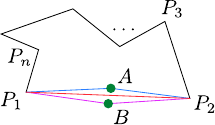
    \caption{Illustration for the inclusion-exclusion principle.}
    \label{fig:explain_polygon_p}
\end{figure}

Let us explain more formally what we mean.
We say a point set is in general position when no three points in the set are collinear.
Let $P = P_1 P_2 \dots P_n$ be an arbitrary polygon with vertices in general position,
depicted in \cref{fig:explain_polygon_p}.
The edge $P_1 P_2$ is marked in red.
Let $A$ be a point very near edge $P_1 P_2$ inside the polygon, but not very near either vertex.
Let $B$ be a point very near $A$, but outside the polygon.
Define polygons $P_A = P_1 A P_2 P_3 \dots P_n$
and $P_B = P_1 B P_2 P_3 \dots P_n$.
We get three slightly-different polygons: $P$, $P_A$, and $P_B$.
Then the proposition claims
\[ T(P_A)+T(P) = T(P_B). \]

\begin{proof}
    Each triangulation of $P_B$ either contains the edge $P_1 P_2$, or it doesn't.

    There is a bijection between the triangulations of $P_B$ that \emph{does} contain edge $P_1 P_2$
    and the triangulations of $P$,
    and there is a bijection between the triangulations of $P_B$ that \emph{does not} contain edge $P_1 P_2$
    and the triangulations of $P_A$.
\end{proof}

\begin{remark}
    \label{remark_location_AB_inclusion_exclusion}
    It matters where $A$ and $B$ is along the edge $P_1 P_2$.
    In the situation depicted in \cref{fig:remark_position_along_segment},
    point $A$ is outside angle $\angle FGH$ but point $B$ is inside,
    which leads to complication because
    triangle $GHB$ does not intersect triangle $GHC$, but triangle $GHA$ does.

\begin{figure}
    \centering
    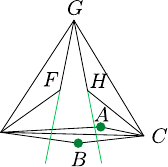
    \caption{Example where the location of $A$ and $B$ along the edge $P_1 P_2$ matters, as pointed out in \cref{remark_location_AB_inclusion_exclusion}.}
    \label{fig:remark_position_along_segment}
\end{figure}

\end{remark}

\subsection{Generalization of Inclusion-Exclusion: Count Triangulations of the Double Circle}

The \emph{double circle} point set is a point set with $2n$ vertices, where $n$ of them forms a regular convex polygon with $n$ vertices, and the remaining $n$ is very near the midpoint of each of the $n$ edges and inside the polygon.

\begin{figure}
    \centering
    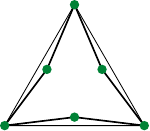
    \caption{Illustration for a double circle with $2 \times 3$ vertices.}
    \label{fig:double_circle_3}
\end{figure}

The double circle with $n = 3$, which is just a triangle with $3$ points inside very near the edges, is shown in \cref{fig:double_circle_3}.

\newsavebox{\doublecirclebox}
\savebox{\doublecirclebox}{\raisebox{-0.1em}{
\begingroup%
  \makeatletter%
  \providecommand\color[2][]{%
    \errmessage{(Inkscape) Color is used for the text in Inkscape, but the package 'color.sty' is not loaded}%
    \renewcommand\color[2][]{}%
  }%
  \providecommand\transparent[1]{%
    \errmessage{(Inkscape) Transparency is used (non-zero) for the text in Inkscape, but the package 'transparent.sty' is not loaded}%
    \renewcommand\transparent[1]{}%
  }%
  \providecommand\rotatebox[2]{#2}%
  \newcommand*\fsize{\dimexpr\f@size pt\relax}%
  \newcommand*\lineheight[1]{\fontsize{\fsize}{#1\fsize}\selectfont}%
  \ifx\svgwidth\undefined%
    \setlength{\unitlength}{10.38392901bp}%
    \ifx\svgscale\undefined%
      \relax%
    \else%
      \setlength{\unitlength}{\unitlength * \real{\svgscale}}%
    \fi%
  \else%
    \setlength{\unitlength}{\svgwidth}%
  \fi%
  \global\let\svgwidth\undefined%
  \global\let\svgscale\undefined%
  \makeatother%
  \begin{picture}(1,0.85384981)%
    \lineheight{1}%
    \setlength\tabcolsep{0pt}%
    \put(0,0){\includegraphics[width=\unitlength,page=1]{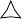}}%
  \end{picture}%
\endgroup%
}}
\newcommand\doublecircle{\usebox{\doublecirclebox}}

How can we compute $T(\doublecircle)$?
The plan is to apply \cref{prop_simple_pie} multiple times:


\begin{equation} \label{equation_double_circle_manipulation}
    \mbox{\raisebox{-1.9em}{
\begingroup%
  \makeatletter%
  \providecommand\color[2][]{%
    \errmessage{(Inkscape) Color is used for the text in Inkscape, but the package 'color.sty' is not loaded}%
    \renewcommand\color[2][]{}%
  }%
  \providecommand\transparent[1]{%
    \errmessage{(Inkscape) Transparency is used (non-zero) for the text in Inkscape, but the package 'transparent.sty' is not loaded}%
    \renewcommand\transparent[1]{}%
  }%
  \providecommand\rotatebox[2]{#2}%
  \newcommand*\fsize{\dimexpr\f@size pt\relax}%
  \newcommand*\lineheight[1]{\fontsize{\fsize}{#1\fsize}\selectfont}%
  \ifx\svgwidth\undefined%
    \setlength{\unitlength}{208.7173233bp}%
    \ifx\svgscale\undefined%
      \relax%
    \else%
      \setlength{\unitlength}{\unitlength * \real{\svgscale}}%
    \fi%
  \else%
    \setlength{\unitlength}{\svgwidth}%
  \fi%
  \global\let\svgwidth\undefined%
  \global\let\svgscale\undefined%
  \makeatother%
  \begin{picture}(1,0.25776472)%
    \lineheight{1}%
    \setlength\tabcolsep{0pt}%
    \put(0,0){\includegraphics[width=\unitlength,page=1]{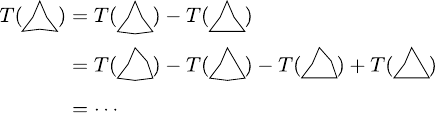}}%
  \end{picture}%
\endgroup%
}}
\end{equation}

In the first step, we apply \cref{prop_simple_pie} on the bottom edge to write $T(\doublecircle)$ as a difference between two $T(-)$ expressions.
After all the expansions, we get
\[ T(\doublecircle) = T(\hexago) - 3 T(\pentago) + 3 T(\square) - T(\triangle). \]

Note that we already know how to compute
$T(\hexago), T(\pentago), T(\square), T(\triangle)$ ---
they are just Catalan numbers.

\subsection{Some Informal Umbral Calculus}

Note that the form of \cref{equation_double_circle_manipulation} is very similar to
the expansion of an expression of the form $(a-b) \cdot (c-d) \cdot (e-f)$:
\begin{align*}
    (a-b)(c-d)(e-f)
    &= a(c-d)(e-f) - b(c-d)(e-f) \\
    &= ac(e-f) - ad(e-f) \\
    & \qquad {} - bc(e-f) + bd(e-f) \\
    &= \cdots
\end{align*}
With some informal abuse of notation, we can write

\[
    [
        \raisebox{-0.5em}{\ 
\begingroup%
  \makeatletter%
  \providecommand\color[2][]{%
    \errmessage{(Inkscape) Color is used for the text in Inkscape, but the package 'color.sty' is not loaded}%
    \renewcommand\color[2][]{}%
  }%
  \providecommand\transparent[1]{%
    \errmessage{(Inkscape) Transparency is used (non-zero) for the text in Inkscape, but the package 'transparent.sty' is not loaded}%
    \renewcommand\transparent[1]{}%
  }%
  \providecommand\rotatebox[2]{#2}%
  \newcommand*\fsize{\dimexpr\f@size pt\relax}%
  \newcommand*\lineheight[1]{\fontsize{\fsize}{#1\fsize}\selectfont}%
  \ifx\svgwidth\undefined%
    \setlength{\unitlength}{39.46060181bp}%
    \ifx\svgscale\undefined%
      \relax%
    \else%
      \setlength{\unitlength}{\unitlength * \real{\svgscale}}%
    \fi%
  \else%
    \setlength{\unitlength}{\svgwidth}%
  \fi%
  \global\let\svgwidth\undefined%
  \global\let\svgscale\undefined%
  \makeatother%
  \begin{picture}(1,0.29998844)%
    \lineheight{1}%
    \setlength\tabcolsep{0pt}%
    \put(0,0){\includegraphics[width=\unitlength,page=1]{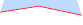}}%
  \end{picture}%
\endgroup%
\ }
    ] \sim [
        \raisebox{-0.5em}{\ 
\begingroup%
  \makeatletter%
  \providecommand\color[2][]{%
    \errmessage{(Inkscape) Color is used for the text in Inkscape, but the package 'color.sty' is not loaded}%
    \renewcommand\color[2][]{}%
  }%
  \providecommand\transparent[1]{%
    \errmessage{(Inkscape) Transparency is used (non-zero) for the text in Inkscape, but the package 'transparent.sty' is not loaded}%
    \renewcommand\transparent[1]{}%
  }%
  \providecommand\rotatebox[2]{#2}%
  \newcommand*\fsize{\dimexpr\f@size pt\relax}%
  \newcommand*\lineheight[1]{\fontsize{\fsize}{#1\fsize}\selectfont}%
  \ifx\svgwidth\undefined%
    \setlength{\unitlength}{39.46060181bp}%
    \ifx\svgscale\undefined%
      \relax%
    \else%
      \setlength{\unitlength}{\unitlength * \real{\svgscale}}%
    \fi%
  \else%
    \setlength{\unitlength}{\svgwidth}%
  \fi%
  \global\let\svgwidth\undefined%
  \global\let\svgscale\undefined%
  \makeatother%
  \begin{picture}(1,0.29998844)%
    \lineheight{1}%
    \setlength\tabcolsep{0pt}%
    \put(0,0){\includegraphics[width=\unitlength,page=1]{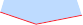}}%
  \end{picture}%
\endgroup%
\ }
    ] - [
        \raisebox{-0.5em}{\ 
\begingroup%
  \makeatletter%
  \providecommand\color[2][]{%
    \errmessage{(Inkscape) Color is used for the text in Inkscape, but the package 'color.sty' is not loaded}%
    \renewcommand\color[2][]{}%
  }%
  \providecommand\transparent[1]{%
    \errmessage{(Inkscape) Transparency is used (non-zero) for the text in Inkscape, but the package 'transparent.sty' is not loaded}%
    \renewcommand\transparent[1]{}%
  }%
  \providecommand\rotatebox[2]{#2}%
  \newcommand*\fsize{\dimexpr\f@size pt\relax}%
  \newcommand*\lineheight[1]{\fontsize{\fsize}{#1\fsize}\selectfont}%
  \ifx\svgwidth\undefined%
    \setlength{\unitlength}{39.46060181bp}%
    \ifx\svgscale\undefined%
      \relax%
    \else%
      \setlength{\unitlength}{\unitlength * \real{\svgscale}}%
    \fi%
  \else%
    \setlength{\unitlength}{\svgwidth}%
  \fi%
  \global\let\svgwidth\undefined%
  \global\let\svgscale\undefined%
  \makeatother%
  \begin{picture}(1,0.29998844)%
    \lineheight{1}%
    \setlength\tabcolsep{0pt}%
    \put(0,0){\includegraphics[width=\unitlength,page=1]{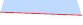}}%
  \end{picture}%
\endgroup%
\ }
    ].
\]

That is:
\begin{moral}
    Whenever
\raisebox{-0.5em}{}
is encountered \emph{in the context of counting number of triangulations above it},
it can be considered as a linear combination
$
[
        \raisebox{-0.5em}{\ \ }
    ] - [
        \raisebox{-0.5em}{\ \ }
    ]$.
\end{moral}
As long as that the middle point is sufficiently close to the edge, that is.

For convenience, we write $x^n$ for the edge consisting of $n$ convex vertices,
and $y^n$ for the edge consisting of $n$ concave vertices.
So,
$x^2 = [
    \raisebox{-0.5em}{\ \ }
]$,
$x^1 = y^1 =  [
    \raisebox{-0.5em}{\ \ }
]$,
and $y^2 = [
    \raisebox{-0.5em}{\ \ }
]$.
The equation above can be written $y^2 \sim x^2-x$.

We abbreviate $x^1$ to $x$, $y^1$ to $y$, and for an integer $a > 0$,
\[ a x^n = \underbrace{x^n + \cdots + x^n}_{a \text{ times}}. \]

\begin{remark}
    Note that ``in the context of counting number of triangulations above it'' is important.
    The formula is reversed when the number of triangulations \emph{below} it is counted instead.
\end{remark}

Now we can see what \cref{equation_double_circle_manipulation} is doing.
We rewrite the concave vertices as a \emph{linear combination} of convex vertices,
and once all vertices are convex, each term is very easy to compute --- again, they're just Catalan numbers.

\subsection{Key Observation of \cite{hurtado1997} and \cite{bacher2010}}
\label{subsection_defining_M}

\cite{bacher2010} does much more than that: it gives an expression of $y^n$ in terms of $x^n$ for every $n$,
which it calls the ``maximal edge polynomials'' $p_n$. In our notation:
\begin{align*}
    y &\sim x \\
    y^2 &\sim x^2-x \\
    y^3 &\sim x^3-2x^2 \\
    y^4 &\sim x^4-3x^3+x^2 \\
    y^5 &\sim x^5 - 4x^4 + 3x^3\\
    y^6 &\sim x^6 - 5x^5 + 6x^4 - x^3\\
    y^7 &\sim x^7 - 6x^6 + 10x^5 - 4x^4\\
    y^8 &\sim x^8 - 7x^7 + 15x^6 - 10x^5 + x^4\\
    y^9 &\sim x^9 - 8x^8 + 21x^7 - 20x^6 + 5x^5
\end{align*}
Morally, the idea is just:
\begin{moral}
    The maximal edge polynomial of a near-edge
    represents the decomposition of that near-edge into a linear combination of convex chains.
\end{moral}

This also appears at \cite[A115139]{oeis}.  A closed-form is also provided in \cite{bacher2010}:
\begin{equation}
    \label{yn_closed_form}
    y^n \sim \sum_{k=0}^{\lfloor n/2 \rfloor} (-1)^k \binom{n-k}{k} x^{n-k}
\end{equation}

For notational convenience, we define a vector space homomorphism
\[ \mathcal M \colon \QQ[y] \to \QQ[x] \]
by $\mathcal M(1) = 1$ and $\mathcal M$ acts on the basis elements in such a way that $y^n \sim \mathcal M(y^n)$,
using the formula in \cref{yn_closed_form}.

We have the following generating function, whose coefficients are the values $\mathcal M(y^i)$:
\begin{equation}
    \label{gen_func_M}
    \frac{1}{1 + (t^2 - t) x} = 
    1 + x t + (x^2 - x) t^2 + (x^3 - 2 x^2) t^3 + \cdots
\end{equation}

\subsection{Relation with \cite{rutschmann2023}}
\label{subsection_defining_T}

\cite{rutschmann2023} defines an \emph{upper triangulation polynomial} of a chain.
Once again, the moral here is:
\begin{moral}
    The upper triangulation polynomial of a near-edge
    represents the decomposition of that near-edge into a linear combination of concave chains.
\end{moral}

Some small modification is needed, in particular the order of coefficients need to be reversed.

The key idea is:
\begin{moral}
    The upper triangulation polynomial is easy to compute with respect to the $\wedge$ operation (just multiply them together), but not with respect to the $\vee$ operation.
    Conversely, the maximal edge polynomial is easy to compute with respect to $\vee$, but not with respect to $\wedge$.
\end{moral}

See \cref{defining_property_homomorphism} for a more formal treatment of this.

This is also very natural: convex near-edge can be easily $\vee$-ed, and concave near-edge can be easily $\wedge$-ed.

In the notation of \cite{rutschmann2023} (we won't define the upper triangulation polynomials as in \cite{rutschmann2023} because we will not use this elsewhere):
\begin{align*}
    T_{\Ccvx(1)}(x) &= 1 \\
    T_{\Ccvx(2)}(x) &= 1+x \\
    T_{\Ccvx(3)}(x) &= 1+2x+2x^2 \\
    T_{\Ccvx(4)}(x) &= 1+3x+5x^2+5x^3
\end{align*}
The last line above is given as an example after \cite[Definition 21]{rutschmann2023}.

In our notation:
\begin{align*}
    x &\sim y \\
    x^2 &\sim y^2+y \\
    x^3 &\sim y^3+2y^2+2y \\
    x^4 &\sim y^4+3y^3+5y^2+5y\\
    x^5 &\sim y^5 + 4y^4 + 9y^3 + 14y^2 + 14y\\
    x^6 &\sim y^6 + 5y^5 + 14y^4 + 28y^3 + 42y^2 + 42y\\
    x^7 &\sim y^7 + 6y^6 + 20y^5 + 48y^4 + 90y^3 + 132y^2 + 132y
\end{align*}

The coefficients are just the entries in Catalan's triangle \cite[A009766]{oeis}.
The closed form is:
\begin{equation} \label{xn_closed_form}
    x^n \sim \sum_{k=1}^n \binom{2n-k}{n-k} \frac{k}{2n-k} y^k.
\end{equation}

Again, for notation convenience we define
\[ \mathcal T \colon \QQ[x] \to \QQ[y] \]
by $\mathcal T(1) = 1$ and $\mathcal T$ acts on the basis elements to satisfy $x^n \sim \mathcal T(x^n)$,
using the closed-form in \cref{xn_closed_form}.

We also have the generating function:
\[
    \frac{2}{2-y+y\sqrt{1-4t}}
    = 1 + y t + (y^2 + y) t^2 + (y^3 + 2 y^2 + 2 y) t^3 + \cdots
\]
Notice the similarity with the generating function of Catalan numbers, where $\sqrt{1-4t}$ term also appear.

\section{Formal Definitions}
\label{sec_formal_definition}

\subsection{Chains and Near-Edges}

In this section, we will formally define chains and near-edges.
Let $P$ be a point set in general position, embedded in the plane $\RR^2$.

We define a chain following \cite{rutschmann2023}.

\begin{definition}[Chain]
    Suppose the points in $P$ has all $x$-coordinates distinct.
    Let $(P_1, P_2, \dots, P_n)$ be all points in $P$, sorted in increasing $x$-order.
    We say $P$ is a chain if for every integer $1 \leq i<n$,
    the edge $P_i P_{i+1}$ is contained in every triangulations of $P$.
\end{definition}
Note that this condition is equivalent to: for every integer $1 \leq i<n$,
the edge $P_i P_{i+1}$ does not intersect any other edge of $P$.

We define a near-edge as follows.
\begin{definition}[Standalone near-edge]
    When all points in $P$ has pairwise distinct $x$-coordinates,
    let $(P_1, P_2, \dots, P_n)$ be all points in $P$ in increasing $x$-order,
    then the polyline $P_1 P_2 \dots P_n$ is called a near-edge.
\end{definition}
This definition is almost tautological, however --- almost every point sets are near-edges that way.
More interestingly, we define the \emph{near-edge inside a point set}.

\begin{definition}[Shrink an object towards a line]
    Let $P$ be a point and $\ell$ be a line.
    For a real number $0 \leq \varepsilon \leq 1$, let $\shrink_\varepsilon(P \to \ell)$
    be a point $P'$ satisfying the following:
    let $H$ be the orthogonal projection of $P$ onto $\ell$,
    then $P'$ lies on line $PH$ and $P' H =\varepsilon \cdot PH$.
    For a polyline $P_1 P_2 \dots P_n$,
    define $\shrink_\varepsilon(P_1 P_2 \dots P_n \to \ell)$
    to be the polyline
    consisting of points $\shrink_\varepsilon(P_i \to \ell)$
    for $1 \leq i \leq n$ in order.
\end{definition}

\begin{definition}[Order type]
    For two ordered tuples of points
    $(A_1, A_2, \dots, A_k)$
    and
    $(A'_1, A'_2, \dots, A'_k)$,
    we say they have the same order type
    if for every $i<j<k$,
    the three points $(A_i, A_j, A_k)$ are in counterclockwise order
    if and only if
    the three points $(A'_i, A'_j, A'_k)$ are in counterclockwise order.

    We define the order type of an (ordered) point set $P$ to be the equivalence
    class of all ordered point sets with the same order type as $P$.
\end{definition}

The concept of order types is very commonly seen in the context of triangulation,
because many combinatorial properties of a point set, such as the number of triangulations,
only depends on its order type.

For real $0<\varepsilon \leq 1$,
line $\ell$,
points $A_1$, $A_2$, $\dots$, $A_k$,
$B_1$, $B_2$, $\dots$, $B_m$,
let point set $P = \{ A_1, \dots, A_k, B_1, \dots, B_m \}$ be in general position,
we say
``shrinking a polyline $A_1 A_2 \dots A_k$ towards $\ell$
by a factor of $\varepsilon$
does not change the order type of $P$''
to mean the following:
let $A'_1 A'_2 \dots A'_k = \shrink_\varepsilon(A_1 A_2 \dots A_k \to \ell)$,
then the ordered point set
$(A_1, \dots, A_k, B_1, \dots, B_m)$
and
$(A'_1, \dots, A'_k, B_1, \dots, B_m)$
have the same order type.
We say
``shrinking a polyline $A_1 A_2 \dots A_k$ towards $\ell$
does not change the order type of $P$''
if the statement above holds for all $0<\varepsilon \leq 1$.

\begin{definition}[Near-edge of a point set]
    Let $P$ be a point set, and $A_1$, $\dots$, $A_k$ be points in $P$
    for integer $k \geq 2$.
    We call the polyline $A_1 A_2 \dots A_k$ a near-edge of the point set $P$
    when there is some rotated coordinate axis with origin $A_1$,
    the points $A_1$, $A_2$, $\dots$, $A_k$ have strictly increasing $x$-coordinate,
    and shrinking $A_1 A_2 \dots A_k$ towards the line $A_1 A_k$
    does not change the order type of $P$.
\end{definition}

We see that our definition is very similar to that in \cite{bacher2010}.

\subsection{Gluing Operation, Convex and Concave Sum}

\begin{definition}[Gluing a near-edge to a point set]
    Let $P$ be a point set in general position.
    Let $M$ and $N$ be two distinct points in $P$.
    Assume there are no two points $F$, $G$ in $P$
    such that line $FG$ intersects the interior of segment $MN$.
    Let $A = A_1 A_2 \dots A_k$ be a standalone near-edge with $k \geq 2$.
    Define the point set $\glue(P, A \to MN)$ obtained by gluing the near-edge $A$ to the edge $MN$ of point set $P$ as follows:
    let $\sigma$ be an orientation-preserving affine transformation
    such that $\sigma(A_1) = M$, $\sigma(A_k) = N$,
    and $\sigma(A_1) \sigma(A_2) \dots \sigma(A_k)$
    is a near-edge of $P \cup \{\sigma(A_2) \dots \sigma(A_k)\}$,
    then.
\end{definition}

For this definition to make sense, we need the following:
\begin{lemma}
    Let $A_1 A_2 \dots A_k$ be a polyline, with $A_1$ have smaller $x$-coordinate than $A_k$.
    Let $B_1$, $\dots$, $B_m$ be points, and let $P = \{ A_1, \dots, A_k, B_1, \dots, B_m \}$.
    Suppose $P$ is in general position.
    Let $\ell$ be the line $A_1 A_k$.
    Then there exists $\varepsilon>0$ such that
    shrinking $A_1 A_2 \dots A_k$ by a factor of $\varepsilon$ towards $\ell$ makes
    it a near-edge of $P$.
    
    More formally, there exists $\varepsilon>0$ such that
    $\shrink_\varepsilon(A_1 A_2 \dots A_k \to \ell)$
    is a near-edge of
    $\shrink_\varepsilon(A_1 A_2 \dots A_k \to \ell) \cup \{ B_1, \dots, B_m \}$.
\end{lemma}
The two lemmas above justifies the existence of $\sigma$ required in the definition above.

We also see that in certain cases, such as when $P$ is the set of vertices of a convex polygon and $MN$ is an edge of the polygon,
the order type of $\glue(P, A \to MN)$ only depends on the order types of $P$ and $A$.

More generally, we may define the point set obtained by gluing multiple near-edges to a point set,
say $\glue(P, A \to M_1 N_1, B \to M_2 N_2)$. Analogous properties hold.

This definition makes it easy to define convex sum and concave sum:
\begin{definition}[Convex sum]
    Let $A = A_1 A_2 \dots A_k$ and $B = B_1 B_2 \dots B_m$ be near-edges.
    Then the near-edge $C = A \vee B$ is defined as follows:
    let $P = \{D, E, F\}$ where $D = (0, 0)$, $E = (1, -1)$, $F = (2, 0)$,
    then let \[ C = \glue(P, A \to DE, B \to EF). \]
\end{definition}

For this definition to make sense, we also need the following.
\begin{lemma}
    Let $A_1 A_2 \dots A_k$ be a polyline, with $A_1$ have smaller $x$-coordinate than $A_k$.
    Let $\ell$ be the line $A_1 A_k$.

    Then there exists $\varepsilon>0$ such that $\shrink_\varepsilon(A_1 A_2 \dots A_k \to \ell)$
    has increasing $x$-coordinate.
\end{lemma}
As mentioned above, while the exact coordinates of the convex sum is not well-defined,
the order of the vertices appearing along the polyline and the order type is.

Concave sum is defined analogously, but the coordinate $E=(1, -1)$ is changed to $(1, 1)$.
We see that our definition of the convex and concave sum is similar to
\cite{rutschmann2023}, except that our definition works for the more general class of near-edges.

Also similar to \cite{rutschmann2023}, we define the flipping operation on near-edge:
given a near-edge $A$, let $\overline A$ be the chain obtained by flipping $A$ vertically.

\section{The Algebraic Theory of Near-Edges}
\label{sec_algebraic_definitions}

As stated at the beginning, we aim to develop a theory to compute the number of near-edges.
Because all point sets are in fact near-edges, this is inherently limited.
Instead, we will do the following:
\begin{itemize}
    \item Define bivariate polynomials associated to each near-edge $A$ that we call the \emph{joint triangulation polynomial}.
    \item Provide algorithms to compute the joint triangulation polynomials of $A \vee B$, $A \wedge B$, $\overline A$
        given the joint triangulation polynomials of $A$ and $B$.
\end{itemize}

\subsection{Univariate Triangulation Polynomials}

Because it is simpler, we will first define univariate polynomials, which are only defined on near-edges that are \emph{chains}.

Following \cite{rutschmann2023}, we define:
\begin{definition}[Primitive chain]
    \label{def_primitive_chain}
    The primitive chain $E$ is the chain consisting of two points $(0, 0)$ and $(1, 0)$.
\end{definition}
\begin{definition}[Convex and concave chain]
    For an integer $i \geq 1$, define
    \[
        \Ccvx(i) = \underbrace{E \vee E \vee \dots \vee E}_{i \text{ copies of } E},\quad
        \Cccv(i) = \underbrace{E \wedge E \wedge \dots \wedge E}_{i \text{ copies of } E}.
    \]
\end{definition}

Recall the vector space isomorphisms $\mathcal T \colon \QQ[x] \to \QQ[y]$ and $\mathcal M \colon \QQ[y] \to \QQ[x]$
defined in \cref{subsection_defining_T} and \cref{subsection_defining_M}. We have:
\begin{lemma}
    $\mathcal T$ and $\mathcal M$ are inverses of each other.
\end{lemma}

\begin{definition}[$t$-polynomial of a chain]
For any chain $C$, we define
\[ t_C (y) = \sum_{i \geq 1} 
    \Big( \text{\begin{varwidth}{15em} number of upper triangulations \\ with $i$ segments \end{varwidth}} \Big)
\cdot y^i. \]
\end{definition}

The relation between the definition above and \cite{rutschmann2023} is the following:
let $T_C$ be the upper triangulation polynomial as defined in \cite{rutschmann2023},
let $n$ be the number of segments in the chain $C$, then
\[ t_C (y) = y^n T_C (1/y). \]
We will not need to use $T_C$ polynomial again in this article.

\begin{definition}[$m$-polynomial of a chain]
    For any chain $C$, we define
    \[ m_C (x) = \mathcal M(t_C(y)). \]
\end{definition}
As such, 
for any two chains $C$ and $D$,
\[ m_{C \vee D}(x) = m_C (x) \cdot m_D (x) \text{ and } 
t_{C \wedge D}(y) = t_C (y) \cdot t_D (y) . \]
Consequently, for each $i\geq 1$,
\[ m_{\Ccvx(i)} (x) = x^i\text{ and } t_{\Cccv(i)} (y) = y^i. \]
This, together with linearity, can also be used as the definition of $t$ and $m$.

In more formal algebraic language:
\begin{prop}
    \label{defining_property_homomorphism}
    Let $\mathcal C$ be the set of all chains. Then:
    \begin{itemize}
        \item the mapping $t_\bullet$ from the monoid $(\mathcal C, \wedge)$ to the monoid
            $(\QQ[y], \cdot)$ is a homomorphism of monoids;
        \item the mapping $m_\bullet$ from the monoid $(\mathcal C, \vee)$ to the monoid
            $(\QQ[x], \cdot)$ is a homomorphism of monoids.
    \end{itemize}
\end{prop}
For short, we will just say $t_\bullet$ is multiplicative over $\wedge$
and $m_\bullet$ is multiplicative over $\vee$.

\begin{definition}[$\vee$ operator]
    \label{polynomial_or_operator}
    Define the binary operator $\vee\colon \QQ[y] \times \QQ[y] \to \QQ[y]$ as follows:
    for any $t_1(y), t_2(y) \in \QQ[y]$,
    \[ t_1(y) \vee t_2(y) = \mathcal T(\mathcal M(t_1(y)) \cdot \mathcal M(t_2(y))). \]
\end{definition}
\begin{definition}[$\wedge$ operator]
    \label{polynomial_and_operator}
    Define the binary operator $\wedge\colon \QQ[x] \times \QQ[x] \to \QQ[x]$ as follows:
    for any $m_1(x), m_2(x) \in \QQ[x]$,
    \[ m_1(x) \wedge m_2(x) = \mathcal M(\mathcal T(m_1(x)) \cdot \mathcal T(m_2(x))). \]
\end{definition}

The motivation for these definitions are clear:
for any two chains $C$ and $D$,
\[
    m_{C \wedge D}(x) = m_C (x) \wedge m_D (x)\text{, }t_{C \vee D}(y) = t_C (y) \vee t_D (y).
\]

\subsubsection{Lower Triangulation}

Similar to \cite{rutschmann2023}, now we consider the lower triangulation.
\begin{definition}[Lower triangulation polynomials]
    For a chain $C$, define $t^*_C (u) \in \QQ[u]$ satisfying $t^*_C (y) = t_{\overline C} (y)$,
    and $m^*_C (v) \in \QQ[v]$ satisfying $m^*_C (v) = m_{\overline C}(x)$.
\end{definition}
Notice that $t^*_C (u)$ is just the lower triangulation polynomial defined in \cite{rutschmann2023},
and $m^*_C = \mathcal M(t^*_C)$. By abuse of notation, we apply $\mathcal M$ on polynomials with different variables here.

Of course, $t^*_\bullet$ is multiplicative over $\vee$ and $m^*_\bullet$ is multiplicative over $\wedge$.

\subsubsection{Joint Triangulation}

\begin{definition}[Joint triangulation polynomials]
    \label{def_chain_joint_triangulation}
    For a chain $C$, define:
    \begin{align*}
        a^{xu}_C (x, u) &= m_C (x) \cdot t^*_C (u) \\
        a^{xv}_C (x, v) &= m_C (x) \cdot m^*_C (v) \\
        a^{yu}_C (y, u) &= t_C (y) \cdot t^*_C (u) \\
        a^{yv}_C (y, v) &= t_C (y) \cdot m^*_C (v).
    \end{align*}
\end{definition}
Each of them are bivariate polynomial in, for example, $\QQ[x, u]$.
Also notice that $\QQ[x, u] = \QQ[x] \otimes \QQ[u]$ as $\QQ$-vector space.

Define $\mathcal M^1\colon \QQ[y] \otimes \QQ[u] \to \QQ[x] \otimes \QQ[u]$ to operate on the first component of the tensor,
similar for $\mathcal M^1 \colon \QQ[y] \otimes \QQ[v] \to \QQ[x] \otimes \QQ[v]$,
also similar for $\mathcal M^2$, $\mathcal T^1$, $\mathcal T^2$.
Then for any chain $C$,
\begin{align*}
    a^{xv}_C &= \mathcal M^2 (a^{xu}_C), \\
    a^{yv}_C &= \mathcal T^1 (a^{xv}_C), \\
             & \quad \vdots
\end{align*}

Evidently $a^{xu}_\bullet$ is multiplicative over $\vee$,
and $a^{yv}_\bullet$ is multiplicative over $\wedge$.

\subsection{Bivariate Triangulation Polynomials}

We generalize the insight in \cite{rutschmann2023} --- consider both upper triangulation polynomial and lower triangulation polynomial --- to near-edges that are not necessarily chains.
In doing that, we will define the joint triangulation polynomials for arbitrary near-edges
satisfying analogous properties to \cref{def_chain_joint_triangulation} ---
in fact, the definitions coincide for chains.

\begin{figure}
    \centering
    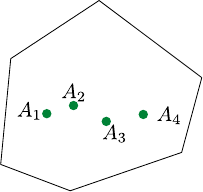
    \caption{Example of a near-edge in a point set.}
    \label{fig:near_edge_with_points_both_sides}
\end{figure}

Consider an arbitrary point set $P$ containing a near-edge $A_1 A_2 \dots A_n$.
See \cref{fig:near_edge_with_points_both_sides} for an illustration.

\begin{figure}
    \centering
    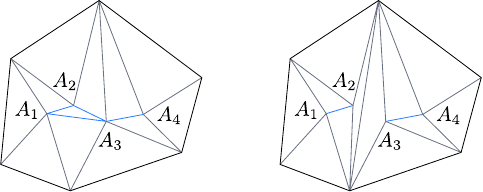
    \caption{Example of a splitting (left) and a nonsplitting (right) triangulation.}
    \label{fig:splitting_and_nonsplitting_triangulation}
\end{figure}

Consider a triangulation $T$ of $P$.
We call $T$ \emph{splitting} with respect to the near-edge $A = A_1 A_2 \dots A_n$
if for every $1 \leq i \leq n-1$, there exists $j \leq i < k$
such that edge $A_j A_k$ belongs to $T$.
Intuitively, this means when the near-edge $A$ is flattened,
the edges belong to the triangulation covers the whole segment $A_1 A_n$,
thus splits the part above $A_1 A_2 \dots A_n$ from the part below it.
See figure~\ref{fig:splitting_and_nonsplitting_triangulation} for an illustration.

Let $T$ be a splitting triangulation as above.
Consider the set of edges and triangles of $T$ that belongs to the near-edge $A$,
and a local coordinate axis such that points $(A_1, \dots, A_n)$ have increasing $x$-coordinate.
Define the \emph{roof} to be the polyline that goes from $A_1$ to $A_n$ through the edges of $T$
with both endpoints belong to $A$, and has the largest $y$-coordinate possible,
and define the \emph{floor} similarly but with the smallest $y$-coordinate possible.

For example, with $T$ being the triangulation depicted
in the left panel of \cref{fig:splitting_and_nonsplitting_triangulation},
the roof is the polyline $A_1 A_2 A_3 A_4$,
and the floor is the polyline $A_1 A_3 A_4$.

\begin{lemma}
    Both the roof and the floor is a polyline that goes through a subset of points $A_i$,
    including $A_1$ and $A_n$, in increasing order.
    Furthermore, all edges in the triangulation $T$ with both endpoints in $A$
    are contained in the region between the roof and the floor.
\end{lemma}
We call a polyline $A_{i_1}A_{i_2}\dots A_{i_p}$ \emph{monotone} if $i_1<i_2<\dots<i_p$.

Now, let $A$ be a standalone near-edge.

\begin{definition}
    \label{def_joint_triangulation}
    Define the joint triangulation polynomial
    \[ a^{yu}_A (y, u) = \sum_T y^{|U|} u^{|L|} \in \QQ[y, u] \]
    where the sum is over all $T$ being a set of edges of $A$ such that:
    \begin{itemize}
        \item there exists monotone polylines $U$ and $L$, both contains $A_1$ and $A_n$;
        \item at any $x$ coordinate between $A_1$ and $A_n$, the polyline $U$ is not below $L$;
        \item all points $A_i$ in $A$ is between $U$ and $L$;
        \item all edges in $U$ and $L$ belong to $T$;
        \item the region between $L$ and $U$ is fully triangulated.
    \end{itemize}
    The last condition can be equivalently expressed as one of the following:
    \begin{itemize}
        \item $T$ is a maximal non-intersecting subset of the
            edges of $A$ that is between $U$ and $L$.
        \item $T$ is a triangulation of the (possibly degenerate) polygon formed by concatenating $U$ and $L$.
    \end{itemize}

    Here, $|U|$ denotes the number of segments in the monotone polyline $U$,
    similar for $|L|$.


    Also define the joint triangulation polynomials
    \begin{align*}
        a^{xu}_A (x, u) &= \mathcal M^1(a^{yu}_A (y, v)), \\
        a^{yv}_A (y, v) &= \mathcal M^2(a^{yu}_A (y, u)), \\
        a^{xv}_A (x, u) &= \mathcal M^1(a^{yv}_A (y, v)).
    \end{align*}
\end{definition}
Then:
\begin{prop}
    \label{joint_triangulation_polynomial_multiplicative}
    $a^{xu}_\bullet$ is multiplicative over $\vee$, and $a^{yv}_\bullet$ is multiplicative over $\wedge$.
\end{prop}

With our definition then $a^{yu}_A (y, u)$ is divisible by $y \cdot u$,
except in a corner case where $A$ only has $1$ point,
in that case it is reasonable to define $a^{yu}_A (y, u) = 1$.

Note that \cref{joint_triangulation_polynomial_multiplicative}
together with closed-form formulas for $\mathcal T$ and $\mathcal M$
gives a polynomial-time algorithm to compute the joint triangulation polynomial
of $A \vee B$ and $A \wedge B$, given the joint triangulation polynomial of $A$ and $B$ ---
thus we have achieved the stated objective at the beginning of this section.

\begin{remark}
    The algorithm for computing joint triangulation polynomials of $\overline A$ given
    joint triangulation polynomials of $A$ is simple --- simply swap the variables,
    equivalently, transpose the coefficient matrix:
    \[
        a^{yu}_{\overline A} (y, u) = a^{yu}_A (u, y).
    \]
\end{remark}

\section{Application: Counting Triangulations}
\label{sec_application_count_triangulations}

We will see that, given the joint triangulation polynomials, it is easy to compute the number of triangulations
of a near-edge.

\begin{prop}
Let $A$ be a near-edge, with its upper hull having $i$ edges and lower hull having $j$ edges.
Then the number of triangulations of $A$ is
\[ 
    [y^i u^j] a^{yu}_A (y, u) .
\]
\end{prop}
Where, following the notation in \cite{Wilf_2005}, we define $[y^i u^j] f(y, u)$ to be the coefficient in $f(y, u)$ corresponding to the monomial $y^i u^j$.

Note that when $A$ is a chain, then $a^{yu}_A (y, u) = t_A (y) \cdot t^*_A (u)$,
thus the value is equal to
\[ [y^i u^j] a^{yu}_A (y, u) 
    = ([y^i] t_A (y)) \cdot ([u^j] t^*_A (u))
\]
which matches the formula in \cite{rutschmann2023}.

We also consider the problem of counting the number of triangulations
of the point set corresponding to an almost-convex polygon ---
that is, a convex polygon with near-edges glued on each of its edge.

\begin{prop}
Let $A_1, \dots, A_n$ be near-edges, and $P = P_1 P_2 \dots P_n$ be a convex polygon,
with vertices listed in counterclockwise order.
Let $j$ be the total number of edges in the lower hulls of all $A_i$.
Then the number of triangulations of
\[
    \glue(P, A_1 \to P_1 P_2, A_2 \to P_2 P_3, \dots, A_n \to P_n P_1)
\]
is
\[
    [y^1 u^j] \mathcal T^1 \Bigl( \frac{1}{x} \cdot \prod_{i=1}^n a^{x, u}_{A_i} (x, u) \Bigr).
\]
\end{prop}

Note that the gluing makes the upper side of $A_i$ points to the interior of $P$.
This matches the formula in \cite{bacher2010}, and can be derived in a similar manner,
noticing $[y^1] \mathcal T(\frac{1}{x}\cdot x^n) = C_{n-2}$.

For $n \geq 3$, when $A_1$, $\dots$, $A_n$ are chains and $A_n = E$ is the primitive chain,
then $\glue(P, \dots)$ has the same order type as $A_1 \vee A_2 \vee \dots \vee A_{n-1}$,
and the above formula gives that the number of triangulations is then equal to
\begin{multline}
    \label{chain_num_triangulations}
    [y^1 u^{n-1}] \mathcal T^1 \Bigl( \prod_{i=1}^{n-1} a^{x, u}_{A_i} (x, u) \Bigr)
    \\ = 
    [y^1 u^{n-1}] a^{y, u}_{A_1 \vee \dots \vee A_{n-1}} (y, u).
\end{multline}
It is evident that the expression on the right is the correct number of triangulations in this case ---
any partial triangulation with $1$ edge on top and exactly $n-1$ edges on bottom
must be a full triangulation.

\section{Algebraic Expression for \texorpdfstring{$\mathcal M$}{M} and \texorpdfstring{$\mathcal T$}{T}}
\label{algebraic_MT}

In this section, we shows the operator $\mathcal M$ and $\mathcal T$ can almost be defined in closed form.

As an application, we show the following formula: let $t(y) \in \QQ[y]$ and $m(x) = \mathcal M(t(y))$, then
\begin{equation}
    \label{relation_m4_t2}
    m(4) = t(2) + 2 t'(2).
\end{equation}
This will have further application in simplifying the statement of \cite[Theorem 30]{rutschmann2023}.

\subsection{A Curious Formula: Extending the \texorpdfstring{$t$}{t} Series}

In \cite[Definition 38]{rutschmann2023}, a rather curious formula is discovered.
For motivation, we consider the following: let
\begin{align*}
    t_1(y) &= y^4+2 y^3+5 y^2 \\
    t_2(y) &= y^7+2 y^6+5 y^5 \\
    t_3(y) &= y^7+2 y^6+5 y^5-8 y^2-28 y-65 \\
    t_4(y) &= y^3+4 y^2+3 y.
\end{align*}
Then
\begin{align*}
    t_1 \vee t_4 &= y^7 + 7 y^6 + 24 y^5 + 58 y^4 + 97 y^3 + 141 y^2 + 141 y \\
    t_2 \vee t_4 &= y^{10} + 7 y^9 + 24 y^8 + 58 y^7 + 97 y^6 + 149 y^5 + 
    \cdots
\end{align*}
Recall that the $\vee$ operator of two polynomials was defined in \cref{polynomial_or_operator}.

Notice that $t_2 = t_1 \cdot y^3$ --- in other words, the initial sequence of coefficients of $t_1$ and $t_2$
are the same, $(1, 2, 5, 0, 0)$.
Also notice that $t_1 \vee t_4$ and $t_2 \vee t_4$ has a few initial coefficients coinciding ---
in particular $(1, 7, 24, 58, 97)$, but no more.

We may try to modify $t_2$ a bit to make the coefficients coincide:
in doing that, we get $t_3$, and
\begin{multline*}
t_3 \vee t_4 = y^{10} + 7 y^9 + 24 y^8 + 58 y^7 + 97 y^6 + 141 y^5 \\
+ 141 y^4 - 251 y^2 - 186 y. 
\end{multline*}
Now \emph{all} of the coefficients $(1, 7, 24, 58, 97, 141, 141, 0)$ of $t_1 \vee t_4$ appear in $t_3 \vee t_4$!

More generally:
\begin{definition}[{Hat operator for $t(y) \in \QQ[y]$}]
    \label{def_hat_y}
    For each polynomial $t(y)$, we define
\[
    \widehat t(y) = t(y) - \frac{1}{y-1} \cdot t\Bigl(\frac{y}{y-1}\Bigr).
\]
\end{definition}
As we mentioned, this is almost the same as the (upper) triangulation generating function in \cite[Definition 38]{rutschmann2023}.

For example,
\begin{multline*}
    \widehat{t_1}(y) = 
    (y-1)^{-5} \cdot (y^9 - 3 y^8 + 5 y^7 - 15 y^6 + 35 y^5 \\
        {} - 49 y^4 + 35 y^3 - 10 y^2) ,
\end{multline*}
and as a Laurent series of $y^{-1}$,
\[ \widehat{t_1}(y) = y^4 + 2y^3 + 5y^2 - 8y^{-1} - 28y^{-2} - 65y^{-3} - 125y^{-4} + \cdots \]
This explains how we obtained the ``extension'' $t_3$ from $t_1$ in the example above.

More interestingly, this ``extended power series'' also satisfy other nice properties:
for any two polynomials $t_1 (y)$ and $t_2 (y)$, we get
\[ \widehat{t_1} \cdot \widehat{t_2} = \widehat {t_1 \vee t_2} \cdot \widehat 1. \]
Where
\[ \widehat 1 = 1-\frac{1}{y-1} = 1 - y^{-1} - y^{-2} - y^{-3} - \cdots \]
is obtained by treating $1$ as a constant polynomial in $y$ and
apply \cref{def_hat_y}.

We notice that both $\frac{\widehat{t_\bullet}}{\widehat 1}$ and $\mathcal M(t_\bullet)$ are multiplicative over $\wedge$,
which leads us to naturally suspect that they are related.
Indeed, because the hat operator is $\QQ$-linear, and by the universal property of polynomial ring, we have
\[ \mathcal M(t(y)) \Bigl(\frac{\widehat y}{\widehat 1}\Bigr) = \frac{\widehat t(y)}{\widehat 1}. \]

Expanding it out, we get: when $t(y) \in \QQ[y]$ and $m(x) = \mathcal M(t(y))$, then
\begin{equation}
    \label{curious_formula}
    m\Bigl(\frac{y^2}{y-1}\Bigr) \cdot \frac{y-2}{y-1}
    = t(y) - \frac{1}{y-1} \cdot t\Bigl(\frac{y}{y-1}\Bigr).
\end{equation}

Differentiate with respect to $y$ once and substitute $y = 2$, we get \cref{relation_m4_t2}.

When $|y-1|=1$, $y = (y-1) \cdot\overline y$, this can be written more symmetrically as
\[
    m(y \cdot \overline y) \cdot(y-\overline y) = y \cdot t(y) - \overline y \cdot t(\overline y).
\]
Since $m$ and $t$ has all real coefficients, this is equivalent to
\[
    m(|y|^2) \cdot \Im (y) = \Im (y \cdot t(y)).
\]

\subsection{Generalization: Extending the \texorpdfstring{$m$}{m} Series}

In a similar manner, we can extend $m(x)$ to $\widehat m(x)$.

\begin{definition}[{Hat operator for $m(x) \in\QQ[x]$}]
    \label{def_hat_x}
    For each polynomial $m(x)$, we define
    \[
        \widehat m(x) = \frac{1}{\sqrt{1-4/x}} \cdot \Bigl[ m(x) \cdot \frac{\sqrt{1-4/x}}{2} \Bigr] + \frac{m(x)}{2}.
    \]
    Where for a Laurent series $f(x^{-1}) \in \QQ((x^{-1}))$, $[f]$ is the polynomial in $\QQ[x]$
    consisting of the terms with nonnegative powers of $x$.
\end{definition}

For example, let $m(x) = x^2+3x$,
then
\[ \widehat m(x) = x^2 + 3 x + 5 x^{-1} + 21 x^{-2} + 81 x^{-3} + 308 x^{-4} + 
\cdots \]

Of course, we also have: for all polynomials $m_1(x), m_2(x) \in \QQ[x]$,
\[ \widehat{m_1}\cdot\widehat{m_2}=\widehat{m_1 \wedge m_2}\cdot\widehat 1. \]
Where $\wedge$ operator for two polynomials was defined in \cref{polynomial_and_operator},
and $\widehat 1 = \frac{1}{2 \sqrt{1-4/x}}+\frac{1}{2}$
is computed by considering $1$ as a polynomial in $x$
and apply \cref{def_hat_x}.
By abuse of notation, $\widehat 1$ in this section is different from that
in the previous section, but no confusion should arise.

Therefore,
\[
    \mathcal T(m(x))\Bigl(\frac{\widehat x}{\widehat 1}\Bigr) = \frac{\widehat m(x)}{\widehat 1}.
\]

\section{Another Analysis for Poly Chains and Twin Chains}
\label{another_analysis_polytwin}

In \cite{rutschmann2023}, poly chains and twin chains were defined.
We can define the analogous concepts for near-edges.
\begin{definition}
    Let $A$ be a near-edge.
    For integers $N \geq 1$, the poly-$A$ near-edge is
    $\Cpoly(A, N) = \underbrace{\overline{A} \vee \dots \vee \overline{A}}_{N\text{ copies}}$.
\end{definition}
\begin{definition}
    Let $A$ be a near-edge.
    For integers $N \geq 1$, the twin-$A$ near-edge is
    $\Ctwin(A, N) = \overline{\Cpoly(A, N)} \vee E \vee \overline{\Cpoly(A, N)}$.
\end{definition}
Recall $E$ was defined in \cref{def_primitive_chain}.
The $E$ in the middle is actually not too useful for the purpose of maximizing the number of triangulations,
we only keep it for consistency with \cite{rutschmann2023}.

Using our notation, we can rewrite \cite[Theorem 30]{rutschmann2023}.
\begin{theorem}
    \label{twin_chain_count}
    Let $A$ be a chain.
    As $N \to \infty$,
    $\Cpoly(A, N)$ has $\widetilde \Theta(m^*_A (4)^N)$ upper triangulations,
    and $\Ctwin(A, N)$ has $\widetilde \Theta(t_A (2)^{2N})$ upper triangulations.
    Therefore, $\Ctwin(A, N)$ has
    $\widetilde \Theta(a^{yv}_A (2, 4)^{2N})$ triangulations.
\end{theorem}

Recall from \cref{relation_m4_t2} that $m^*_A(4) = t^*_A(2) + 2 t^{*\prime}_A(2)$.

We can give a heuristic argument for the derivation of the first part, the second part is similar.
Similar to \cref{chain_num_triangulations}, when $A$ is a chain and $N\geq 2$,
the number of upper triangulations of $\Cpoly(A, N)$ is
$[y^1] \mathcal T\big(m^*_A (x)^N\big)$.
Write $m^*_A (x)^n = \sum_{i \geq 0} m_i x^i$, then
$[y^1] \mathcal T\big(m^*_A (x)^N\big) = \sum_{i \geq 0} m_i C_{i-2}$.
We have $C_i \in \widetilde \Theta(4^i)$,
so we heuristically expect
$\sum_{i \geq 0} m_i C_{i-2} \approx \sum_{i \geq 0} m_i 4^i$.

In short, the heuristic is, for all chains $A$ whose upper convex hull has length $1$, then
\begin{equation}
    \label{heuristic_m4}
    [y^1] \mathcal T(m_A (x)) \approx m_A (4).
\end{equation}
The approximation is up to a polynomial factor in $n$, the number of vertices in the chain $A$.

Using this heuristic, the number of upper triangulations of $\Cpoly(A, N)$ is $\approx m^*_A (4)^N$
which is exactly what we want.
Unfortunately, we are unable to prove \cref{heuristic_m4}.

\begin{remark}
Of course, \cref{heuristic_m4} cannot be true for all polynomials $m(x) \in \QQ[x]$, the reason is the following.
Let $\mathcal T(m(x)) = t(y) = \sum_{i \geq 0} t_i y^i$,
then the left hand side is equal to $[y^1] \mathcal T(m(x)) = t_1$,
while the right hand side is equal to $t(2) + 2 t'(2)$ which depends on all the coefficients $t_i$.
As such, if for example $t_2$ is very large compared to $t_1$ then \cref{heuristic_m4} does not hold.

When $t(y) = t_A (y)$ for some actual near-edge $A$, we have all $t_i$ are $\geq 0$.
\end{remark}

\section{Numerical Experiments}
\label{sec_numerical_experiment}

\subsection{Finding Point Sets with Many Triangulations}
\label{sec_numerical_experiment_many_triangulations}

We conjecture a generalization of \cref{twin_chain_count} to the case where $A$
is any near-edge instead of just a chain.
\begin{conj}
    \label{conjecture_near_edge_twin_asymptotic}
    Let $A$ be a near-edge. Then
    $\Ctwin(A, N)$ has $\widetilde \Theta(a^{yv}_A (2, 4)^{2N})$ triangulations
    as $N \to \infty$.
\end{conj}

Using this conjecture,
in order to try to improve the bound on
the maximum number of triangulations,
we run the following experiment:
\begin{itemize}
    \item Iterate over all near-edges $A$ with a certain number of vertices.
    \item Use a modified variant of \cite{alvarez2013}
        to compute the joint triangulation polynomial
        of a near-edge with $n$ points in time complexity $\widetilde O(2^n)$.
    \item Compute the joint triangulation polynomial of the near-edge $K_s (A)$ for some integer $s$.
    \item Use \cref{conjecture_near_edge_twin_asymptotic}
        to count the triangulations
        of $\Ctwin(K_s (A), N)$ as $N \to \infty$.
\end{itemize}

Where we define the generalized Koch near-edge as follows:
\begin{definition}
    Let $A$ be a near-edge.
    For integers $s \geq 0$, define
    $K_0 (A) = A$
    and $K_s (A) = \overline{K_{s-1}(A)} \vee \overline{K_{s-1}(A)}$
    for $s \geq 1$.
\end{definition}
When $A = E$ is the primitive chain, $K_s (E)$ is exactly the Koch chain
as defined in \cite{rutschmann2023}.

In order to iterate over all near-edges with $n$ vertices,
notice that for near-edges $A = A_1 A_2 \dots A_n$
and $B = B_1 B_2 \dots B_n$ with the same order type,
then when we add a point $Y$ with very large $y$-coordinate,
$A \cup \{Y\}$ and $B \cup \{Y\}$ has the same order type as well,
because the counterclockwise order of the vertices in $A$ and $B$
is equal to the increasing $x$-coordinate ordering of the vertices on the polyline.
As such, we iterate over all order types with $n+1$ vertices
using the database provided by \cite{Aichholzer_2001},
pick a point $Y$ on the convex hull,
and let the polyline be the points in counterclockwise ordering around $Y$.

Note that \cite{Aichholzer_2001} considers two order types that differ by a reflection
as equivalent, however this is not a problem
because horizontal reflection does not change the triangulation polynomials.

The resulting data is included in \cref{sec_numerical_experiment_result_a}.

\subsection{Rate of Growth for Coefficients of \texorpdfstring{$t_A (y)$}{t\_A (y)}}
\label{sec_coefficient_growth}

As we mentioned earlier, in order for \cref{heuristic_m4} to hold,
we want that for every chain $A$ with upper hull have length $1$,
then $[y^i] t_A (y)$ is not too large compared to $[y^1] t_A (y)$.

We may hope to generalize \cref{heuristic_m4} as follows.

Let near-edge $A$, roof $U$ and floor $L$ as in \cref{def_joint_triangulation}.
Fix one such floor $L$, define $t^L_A (y) = \sum_T y^{|U|}$
where the sum is taken over all triangulations $T$ with floor $L$ and roof $U$.

Since $a^{yu}_A (y, u) = \sum_L t^L_A (y) u^{|L|}$, we may hope the following is true:
for every near-edge $A$, floor $L$,
let $i$ be the number of segments of the convex hull of $A$ and $j > i$,
then $[y^j] t^L_A (y)$ is not too large compared to $[y^i] t^L_A (y)$.

The generalization above is unfortunately false.

\begin{figure}
    \centering
    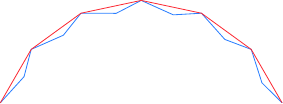
    \caption{The chain $A = \overline{\Cpoly(\Ccvx(2), 6)}$ with $[y^9] t^L_A (y) \gg [y^6] t^L_A (y)$.}
    \label{fig:double_circle_based_chain}
\end{figure}

Consider the chain $A = \overline{\Cpoly(\Ccvx(2), 6)}$ which is similar to part of the double circle,
depicted in \cref{fig:double_circle_based_chain}.
The floor $L$ is fixed (drawn in blue).
Then $[y^6] t^L_A (y) = 1$, because when $|U| = 6$
there can only be one triangulation between $L$ and $U$,
but $[y^9] t^L_A (y) =\binom{6}{3}$ because there are that many choices for $U$.
This example can be generalized by replacing $6$ with any larger even number.

Still, we investigate what is the maximum value of $\frac{[y^i] t^L_A (y)}{[y^j] t^L_A (y)}$ for each $(n, i, j)$ combination.
The resulting data is included in \cref{sec_numerical_experiment_result_coefficient_ratio}.

\subsection{Algorithmic Considerations}

We observe that the $\wedge$ and $\vee$ operator can easily be generalized to polynomials over a finite field.

Over a finite field with sufficiently many roots of unity, the number-theoretic transform
\cite{Agarwal_1975,Pollard_1971},
which is a generalization of the discrete Fourier transform
\cite{Cooley_1965},
can be computed in $O(n \log n)$, thus allowing polynomial multiplication to be computed
in $O(n \log n)$.

On the other hand, it is not as apparent that the $\vee$ and $\wedge$ operator
in \cref{polynomial_or_operator,polynomial_and_operator}
can be computed in $O(n \log n)$, even over a finite field.
One attempt to do so is to use the $\widehat{\bullet}$ operator
in \cref{def_hat_y,def_hat_x}, which allows computing $\widehat m(x)$ from
$m(x)$ successfully, but computing $\widehat t(x)$ from $t(x)$
requires the computation of $t(\frac{y}{y-1})$ from $t(y)$.
With some algebraic manipulation, we see that the problem is no harder than
polynomial multiplication and the problem of computing coefficients of $t(y+1)$
from that of $t(y)$.

Strangely, using the flipped coefficient ordering as in \cite{rutschmann2023}
gives a much nicer transformation (linear shift in the domain)
than $t(y) \mapsto t(\frac{y}{y-1})$.

Computing the coefficients of $t(y+1)$ from the coefficients of $t(y)$ is seen to be equivalent to evaluating $t(1)$, $t'(1)$, $t''(1)$, $\dots$
until the derivative is zero, assume the characteristic is larger than the degree of the polynomial.
The best known algorithm to do so
reduces it to polynomial multipoint evaluation
and interpolation, which takes $O(n \log^2 n)$
\cite{Borodin_1974,kung1973fast,vonzurGathen_Gerhard_2013}.

Over $\QQ$ however, the size of the coefficients appears to grow linearly,
which means there cannot be any $\widetilde O(n)$ algorithm.
Nonetheless, as noted in \cite{rutschmann2023}, it is possible to compute
the coefficients to a fixed relative precision in $\widetilde O(n^2)$ ---
there is an $O(n^2)$ algorithm that computes $t_1(y) \vee t_2(y)$
using only addition and multiplication,
and the values being operated on are nonnegative as long as all coefficients
of $t_1$ and $t_2$ are nonnegative, therefore the rounding errors grows at most linearly.

Noting that this is still worse than the algorithm over finite field,
we hope to find an algorithm that computes either $\vee$ or $\wedge$
in $\widetilde O(n)$ over floating point numbers such that the rounding error
remains controllable.
One idea would be to use multipoint evaluation to evaluate $t$ and/or $m$
at certain points and apply \cref{curious_formula},
however even a single polynomial multiplication can be costly,
noting that polynomial multiplication over floating-point is
at least as hard as convolution in the tropical semiring
\cite{Bateni_2018,Cygan_2019,bringmann_2023}.
The polynomials being considered in this problem is close to convex, which should allow for more efficient algorithms for polynomial multiplication by adapting \cite{bringmann_2023}.
However, when we try to compute $\wedge$ or $\vee$,
the error introduced by rounding intermediate values remains out of control.

\section{Conclusion}

We provide an intuitive interpretation of the algebraic manipulations on generating functions
in \cite{rutschmann2023}. Using that interpretation, we generalize the triangulation polynomials from the class of chains
to the class of near-edges, and show how can the number of triangulations of some classes of almost-convex polygons
can be computed efficiently.

{
    \hfuzz=10pt
\bibliographystyle{abbrv}
\bibliography{article}

\begin{thebibliography}{10}

\bibitem{Agarwal_1975}
R.~Agarwal and C.~Burrus.
\newblock Number theoretic transforms to implement fast digital convolution.
\newblock {\em Proceedings of the IEEE}, 63(4):550–560, 1975.

\bibitem{Aichholzer_2001}
O.~Aichholzer, F.~Aurenhammer, and H.~Krasser.
\newblock Enumerating order types for small sets with applications.
\newblock In {\em Proceedings of the seventeenth annual symposium on Computational geometry}, SoCG01, page 11–18. ACM, June 2001.

\bibitem{aichholzer2007}
O.~Aichholzer, T.~Hackl, C.~Huemer, F.~Hurtado, H.~Krasser, and B.~Vogtenhuber.
\newblock On the number of plane geometric graphs.
\newblock {\em Graphs Comb.}, 23(Supplement-1):67--84, 2007.

\bibitem{alvarez2013}
V.~Alvarez and R.~Seidel.
\newblock A simple aggregative algorithm for counting triangulations of planar point sets and related problems.
\newblock In {\em Proceedings of the 29th the Symposium on Computational Geometry}, 2013.

\bibitem{bacher2010}
R.~Bacher and F.~Mouton.
\newblock Triangulations of nearly convex polygons, 2010.

\bibitem{Bateni_2018}
M.~Bateni, M.~Hajiaghayi, S.~Seddighin, and C.~Stein.
\newblock Fast algorithms for knapsack via convolution and prediction.
\newblock In {\em Proceedings of the 50th Annual ACM SIGACT Symposium on Theory of Computing}, STOC ’18, page 1269–1282. ACM, June 2018.

\bibitem{Borodin_1974}
A.~Borodin and R.~Moenck.
\newblock Fast modular transforms.
\newblock {\em Journal of Computer and System Sciences}, 8(3):366–386, June 1974.

\bibitem{bringmann_2023}
K.~Bringmann and A.~Cassis.
\newblock {Faster 0-1-Knapsack via Near-Convex Min-Plus-Convolution}.
\newblock In I.~L. G{\o}rtz, M.~Farach-Colton, S.~J. Puglisi, and G.~Herman, editors, {\em 31st Annual European Symposium on Algorithms (ESA 2023)}, volume 274 of {\em Leibniz International Proceedings in Informatics (LIPIcs)}, pages 24:1--24:16, Dagstuhl, Germany, 2023. Schloss Dagstuhl -- Leibniz-Zentrum f{\"u}r Informatik.

\bibitem{Cooley_1965}
J.~W. Cooley and J.~W. Tukey.
\newblock An algorithm for the machine calculation of complex fourier series.
\newblock {\em Mathematics of Computation}, 19(90):297, Apr. 1965.

\bibitem{Cygan_2019}
M.~Cygan, M.~Mucha, K.~Węgrzycki, and M.~Włodarczyk.
\newblock On problems equivalent to (min,+)-convolution.
\newblock {\em ACM Transactions on Algorithms}, 15(1):1–25, Jan. 2019.

\bibitem{denny1997}
M.~Denny and C.~Sohler.
\newblock Encoding a triangulation as a permutation of its point set.
\newblock In {\em Proceedings of the 9th Canadian Conference on Computational Geometry}, 1997.

\bibitem{dumitrescu2013}
A.~Dumitrescu, A.~Schulz, A.~Sheffer, and C.~D. T{\'{o}}th.
\newblock Bounds on the maximum multiplicity of some common geometric graphs.
\newblock {\em {SIAM} J. Discret. Math.}, 27(2):802--826, 2013.

\bibitem{hurtado1997}
F.~Hurtado and M.~Noy.
\newblock Counting triangulations of almost-convex polygons.
\newblock {\em Ars Comb.}, 45:169--179, 1997.

\bibitem{kung1973fast}
H.-T. Kung.
\newblock {\em Fast evaluation and interpolation}.
\newblock Carnegie-Mellon University. Department of Computer Science, 1973.

\bibitem{marx2016}
D.~Marx and T.~Miltzow.
\newblock Peeling and nibbling the cactus: Subexponential-time algorithms for counting triangulations and related problems.
\newblock In {\em Proceedings of the 32nd International Symposium on Computational Geometry}, 2016.

\bibitem{oeis}
{OEIS Foundation Inc.}
\newblock The {O}n-{L}ine {E}ncyclopedia of {I}nteger {S}equences, 2025.
\newblock Published electronically at \url{http://oeis.org}.

\bibitem{garcia2000}
A.~G. Olaverri, M.~Noy, and J.~Tejel.
\newblock Lower bounds on the number of crossing-free subgraphs of \(k_n\).
\newblock {\em Comput. Geom.}, 16(4):211--221, 2000.

\bibitem{Pollard_1971}
J.~M. Pollard.
\newblock The fast fourier transform in a finite field.
\newblock {\em Mathematics of Computation}, 25(114):365, Apr. 1971.

\bibitem{rutschmann2023}
D.~Rutschmann and M.~Wettstein.
\newblock Chains, koch chains, and point sets with many triangulations, 2023.

\bibitem{santos2003}
F.~Santos and R.~Seidel.
\newblock A better upper bound on the number of triangulations of a planar point set.
\newblock {\em J. Comb. Theory, Ser. {A}}, 102(1):186--193, 2003.

\bibitem{seidel1998}
R.~Seidel.
\newblock On the number of triangulations of planar point sets.
\newblock {\em Comb.}, 18(2):297--299, 1998.

\bibitem{sharir2011}
M.~Sharir and A.~Sheffer.
\newblock Counting triangulations of planar point sets.
\newblock {\em Electron. J. Comb.}, 18(1):1--74, 2011.

\bibitem{sharir2006}
M.~Sharir and E.~Welzl.
\newblock Random triangulations of planar point sets.
\newblock In {\em Proceedings of the 22nd {ACM} Symposium on Computational Geometry}, 2006.

\bibitem{smith1989}
W.~D. Smith.
\newblock {\em Studies in Computational Geometry Motivated by Mesh Generation}.
\newblock PhD thesis, Princeton University, Princeton, USA, 1989.

\bibitem{vonzurGathen_Gerhard_2013}
J.~von~zur Gathen and J.~Gerhard.
\newblock {\em Fast polynomial evaluation and interpolation}, page 295–312.
\newblock Cambridge University Press, 2013.

\bibitem{Wilf_2005}
H.~S. Wilf.
\newblock generatingfunctionology: Third edition, Dec. 2005.

\end{thebibliography}

}

\clearpage

\appendix
\crefalias{section}{appendix}
\crefalias{subsection}{appendix}

\section{Numerical Experiments Result}
\label{sec_numerical_experiment_result}
\subsection{Finding Point Sets with Large Number of Triangulations}
\label{sec_numerical_experiment_result_a}

We perform the procedure described in 
\cref{sec_numerical_experiment_many_triangulations}
with $n = 9$ (thus the Koch chain
$K_3 = K_3(E)$ is one of the near-edge considered).
For each near-edge $A$ obtained from a point set with $n+1=10$ points,
we compute the joint triangulation polynomial of $K_2 (A)$,
then use \cref{conjecture_near_edge_twin_asymptotic} to compute the rate of growth
of $T(\Ctwin(K_2 (A), N))$ as $N \to \infty$.

None of the near-edge found has number of triangulations of $\Ctwin$ exceed that of the Koch chain.
The raw data obtained for the $30$ point sets with the best asymptotic growth rate is listed here.
\begin{gather*}
(9.02446, 2374662, 3), \\
(9.00649, 2084503, 2), \\
(9.00389, 1253304, 3), \\
(9.00083, 2374664, 1), \\
(9.00068, 1251552, 1), \\
(8.99996, 2084507, 1), \\
(8.99903, 2377696, 3), \\
(8.99434, 2356623, 2), \\
(8.99366, 2084505, 3), \\
(8.99355, 1253337, 3), \\
(8.99297, 2356095, 2), \\
(8.99262, 2377672, 3), \\
(8.99220,  657268, 2), \\
(8.99216, 2356097, 1), \\
(8.99158, 2235342, 1), \\
(8.99093, 1253305, 1), \\
(8.99076, 2356624, 3), \\
(8.99071, 2234551, 2), \\
(8.98998, 2377286, 1), \\
(8.98994, 1240250, 1), \\
(8.98918, 1986810, 3), \\
(8.98916, 1986819, 3), \\
(8.98890, 2375183, 3), \\
(8.98887, 1253346, 3), \\
(8.98851, 1241672, 3), \\
(8.98829, 2377546, 3), \\
(8.98819, 1253277, 3), \\
(8.98785, 1251553, 2), \\
(8.98778, 1252800, 3), \\
(8.98768, 1241675, 2).
\end{gather*}
Each line representing such a near-edge $A$.
The first number is the rate of growth of the number of triangulations,
that is $T(\Ctwin(K_2(A), N)) \in \widetilde \Theta(b^{m})$,
where $m$ is the number of points of $\Ctwin(K_2(A), N)$,
assuming \cref{conjecture_near_edge_twin_asymptotic}.
The other two numbers $(i, j)$ identifies the near-edge $A$ by the following procedure:
in the file \texttt{otypes10.b16} in the database provided in \cite{Aichholzer_2001},
take the $i$-th point set ($0$-indexed),
let its convex hull in counterclockwise order
be $P_0 P_1 \dots P_{k-1}$ where $P_0$ is the point with coordinates having smallest lexicographical order,
then pick $Y$ to be the point $P_j$.

The largest entry, $(9.02446, 2374662, 3)$ corresponds to the Koch chain ---
that is, when $(i, j) = (2374662, 3)$,
$A = K_3$, so $\Ctwin(K_2 (A), N) = \Ctwin(K_5, N)$.

\subsection{Comparing Number of Partial Triangulations per Number of Roof Segments}
\label{sec_numerical_experiment_result_coefficient_ratio}

We provide the result of the numerical experiment described in \cref{sec_coefficient_growth}.
Again we let $n = 9$ and consider all near-edge $A$ with $9$.

{
    \scriptsize

\begin{verbatim}
[(
   [     1      1    1/2    1/5    1/6    1/6    1/4      1]
   [     1      1    1/2    1/5    1/6    1/6    1/4      1]
   [     2      2      1    2/5    1/4    1/4    1/3      1]
   [     3      3      3      1    1/2    1/3    1/3      1]
   [     8      8      7      4      1    1/2    1/2      1]
   [    16     16   31/2     12   50/9      1    1/2      1]
   [ 429/7  429/7 663/11     45 273/10   53/5      1      1]
1, [   660    660    687    577    363    166     39      1]
),
 (
   [     1    1/2    1/5    1/6    1/6    1/4      1]
   [     2      1    2/5    1/4    1/4    1/3      1]
   [     5      3      1    1/2    1/3    1/3      1]
   [    10      9   14/3      1    1/2    1/2      1]
   [    19     19     14   22/3      1    1/2      1]
   [ 429/7 708/11  284/5  177/5 133/10      1      1]
2, [   574    828    746    528    268     69      1]
),
 (
   [     1    1/3    1/5    1/6    1/4      1]
   [   7/2      1    1/2    1/3    1/3      1]
   [    14   11/2      1    1/2    1/2      1]
   [    23   33/2   25/3      1    1/2      1]
   [ 429/7 527/10 381/10   71/5      1      1]
3, [   523    652    560    284     73      1]
),
 (
   [     1    1/4    1/6    1/4      1]
   [  19/4      1    1/2    1/2      1]
   [    21 101/13      1    1/2      1]
   [ 429/8  162/5   61/5      1      1]
4, [   488    482    244     55      1]
),
 (
   [     1    1/3    1/3      1]
   [  33/5      1    1/2      1]
   [    33 329/34      1      1]
5, [   435    171     42      1]
),
 (
   [     1    1/2      1]
   [227/28      1      1]
6, [ 429/4  219/8      1]
),
 (
   [     1      1]
7, [429/25      1]
),
 (8, [1])]
\end{verbatim}

}

In each tuple, the first number is the number of segments in the upper convex hull ---
note that if $A$ is fixed, a floor $L$ is fixed, and its upper hull has $k$ edges,
then the first nonzero coefficient of $t^L_A (y)$ is $[y^k]t^L_A (y)$.
For a fixed $k$, for each $k \leq i \leq j$,
the entry on $i-k+1$-th row and $j-k+1$-th column is the minimum value of
$\frac{[y^i] t^L_A (y)}{[y^j] t^L_A (y)}$,
while the value opposite to it through the main diagonal is the maximum value of the fraction.

\section{On Generating Function of Vector Space Endomorphism}

It is also possible to obtain \cref{curious_formula} as follows:
By partial fraction decomposition of the generating function in \cref{gen_func_M}
\begin{multline*}
    \frac{1}{1+(t^2-t) x} = 
\frac{x+\sqrt{x^2-4x}}{2    \sqrt{x^2-4x} }
\cdot \frac{1}{1-\frac{x+\sqrt{x^2-4x}}{2} t}
\\ {} + \frac{x-\sqrt{x^2-4x}}{2   \sqrt{x^2-4x} }
\cdot \frac{1}{1- \frac{x-\sqrt{x^2-4x}}{2} t}
\end{multline*}
we get
\begin{multline*}
    m(x) = \frac{x+\sqrt{x^2-4x}}{2    \sqrt{x^2-4x} }\, t\Bigl(\frac{x+\sqrt{x^2-4x}}{2}\Bigr)  \\ {} -
    \frac{x-\sqrt{x^2-4x}}{2   \sqrt{x^2-4x} }\, t\Bigl(\frac{x-\sqrt{x^2-4x}}{2}\Bigr).
\end{multline*}
Setting $z = \frac{x+\sqrt{x^2-4x}}{2}$ gives $x = \frac{z^2}{z-1}$,
which allows recovering the original formula.

Note that the transformation $\mathcal T \colon m(x) \mapsto t(y)$ and vice versa
is a linear transformation in the $\QQ$-vector space $\QQ[x]$,
in particular they're invertible endomorphisms.

An endomorphism $\varphi \colon \QQ[x] \to \QQ[y]$
can be represented by a collection
\[ (\varphi_0(y), \varphi_1(y), \dots) \]
which determines where each basis vector $(x^0, x^1, \dots)$
get sent to.

An infinite sequence of elements $(\varphi_0, \varphi_1, \dots)$ in any field $F$ can be represented by a formal power series $\sum \varphi_i t^i \in F[[t]]$.

We would like to note that the two generating functions
\[ \frac{1}{1 + (t^2 - t) x} \]
and
\[ \frac{2}{2-(1-\sqrt{1-4t}) \cdot y} \]
listed above represents the endomorphisms that sends $m(x) \mapsto t(y)$ and vice versa.

Notice also that the elements in $\QQ[-][[t]]$
are \emph{algebraic} over $\Frac(\QQ[-, t])$, which we find quite curious.

The endomorphism
\[ t \mapsto t^{(d)}(p) \]
is represented by the formal power series
\[ \frac{d! \cdot t^d}{(1-p \cdot t)^{d+1}}. \]
In particular the endomorphism $t \mapsto t(p)$
is represented by $\frac{1}{1-p \cdot t}$.
By partial fraction decomposition,
if the base field is algebraically closed
(e.g. replace $\QQ[x]$ with $\CC$ or the algebraic closure of $\QQ[x]$?),
we conclude that an endomorphism
represented with a rational function
can be written as a finite sum of
elements of the form $t \mapsto t^{(d)}(p)$,
thus can be simply written.

When it's not algebraically closed, we imagine that some algebraic geometry is useful here e.g. function evaluation is substituted by taking the function modulo a fixed-degree polynomial.

We also conjecture that the composition (as $\QQ[x]$ endomorphism) of two rational functions is a rational function,
the composition of two elements algebraic over $\QQ[x, t]$
is algebraic over $\QQ[x, t]$,
but this seems difficult to prove, not the least because
we have no idea how to compose two functions represented as elements
in the first place.

Also the identity endomorphism is represented by $\frac{1}{1-xt}$.

\end{document}